\documentclass[useAMS,usenatbib]{mn2e}
\usepackage[english]{babel}
\usepackage[latin1]{inputenc}
\usepackage{amsmath}
\usepackage[dvips]{graphicx}
\newcommand{\sradi}{R_{\odot}}

\title[Kinematic solar dynamo models with a deep meridional flow]{Kinematic solar dynamo models with a deep meridional flow}
\author[G. A. Guerrero and J. D. Muñoz]{G. A. Guerrero$^{1}$\thanks{E-mail:
gaguerreroe@unal.edu.co (GAG); jdmunozc@unal.edu.co
(JDM)} and J. D. 
Muñoz$^{1}$\footnotemark[1]\\
$^{1}$Physics Departament, Universidad Nacional de Colombia, Ciudad Universitaria, Bogotá, 1, Colombia}

\begin{document}
\pagerange{\pageref{firstpage}--\pageref{lastpage}} \pubyear{2003}

\maketitle
\label{firstpage}
\begin{abstract}
We develop two different solar dynamo models to verify the hypothesis that a
deep meridional flow can restrict the apperance of sunspots below 45 
degrees, proposed by \cite{nandi2002}.
In the first one, a single polytropic approximation for the density
profile was  taken, for both radiative and convective zones. 
In the second one, two polytropes were used to distinguish between
both zones  \citep{gpinzon}.
The magnetic buoyancy mechanism proposed by \cite{dikchar99} was
chosen in both models. We, actually, have obtained that 
a deep meridional flow pushes the maxima of toroidal magnetic field
toward the solar equator, but in contrast to \cite{nandi2002}, a second
zone of maximal fields remains at the poles. The second model,
although closely resembling the  solar standard model of
\cite{bahcall,bahcall01}, gives solar cyles three times longer than
observed.   
\end{abstract}

\begin{keywords}
MHD - magnetic field - Sun: interior - Sun: magnetic fields.
\end{keywords}

\section{Introduction}
Despite its irregular appearance, the solar magnetic cycle evolves in a
spatially and temporally well organized manner. Some sunspots
associated phenomena, such as the 11 years cycle of alternating
polarities (Hale law), allows to conclude that the solar magnetic
cycle can be explained by a dynamo 
process involving the transformation of a poloidal magnetic field into
a toroidal one, and the later regeneration of the poloidal field, but of
opposed polarity to the initial, and so on. In the
Babcock-Leighton approach \citep{babcock,leighton} the dynamo operates
in the following way:  
In a first stage, an initial dipolar field, with lines on meridional
planes, is dragged in the east-west direction by the action of the
solar  differential rotation to form a toroidal field.
This happens close to the base of the solar convection zone (SCZ), in
a thin layer called tachocline, where helioseismology has discovered a
substancial radial shear in the rotation pattern. 
In a second stage, tubes of toroidal flux emerge to
the surface, since these tubes of intense 
magnetic field are less dense than their sourrondings (magnetic
bouyancy force). 
While the force lines
rises they are twisted by the Coriolis force to form the bipolar
magnetic regions (BMR) associated with sunspots in the solar
surface.  In the last stage, the leader portion of each BMR (i.e, that
is farther ahead in the direction of solar rotation) migrates
equatorward, while the follower portion migrates poleward, thanks to the
meridional 
circulation. The general effect is that the force lines between the
two BMRs lay on the poloidal direction, but oppossite to the original
field. On the long run, they cancel out the original dipolar
field and give rise to a new one in the opposite direction
\citep{choudhuri95,d95,d96,d97,dikchar99}.  

In the kinematic regime, where a fixed velocity field is assumed
and one investigates just the evolution of the magnetic field due to
that velocitiy field 
(without the back-reaction of the magnetic field onto the solar
plasma) \citep{choudhuri00},
three main ingredients are used to resemble the mass transport
in the Babcock-Leighton approach: differential
rotation, meridional circulation and magnetic buoyancy.  The first one
can be suitably determined from helioseismology measurements, and
resembles closely the radial shear at the tachocline
and the latitudinal angular velocity distribution in the convection
zone. 
The second one is the main flux transport agent. It is responsible to lead
the poloidal flux generated in the surface to the deeper layers
where toroidal flux is regenerated. Obeservational evidence
tell us that there is a poleward meridional flow at the surface with an
average speed between $10$$-$$20$ m s$^{-1}$, but the internal return flow
remains unknown \citep{giles97}. The last ingredient (magnetic bouyancy)
summarizes the 
results of rising magnetic tube simulations
\citep{dsilva93,fan93,caligari95,calig98}, where a buoyant 
magnetic force appears, thanks to the density gradient
between the inside and the outside of the tube. We will incorporate a
simplified form for magnetic buoyancy which resembles an average number
of such events.

Most of the recent kinematic models are able to sucessfully explain the
phenomena like the reversal of the solar dipolar field in an 11-years
cycle or the reversal of polarities of sunspot pairs from one cycle
to the next (Hale law) \citep{dikpati94,dikchar99}. Neverthless, they
fail to predict the absence of sunspots at latitudes above 30 degrees
(Spörer Law). Based on recent numerical simulations which suggest that
a subadiabatic stratification, like the 
one in the solar radiative zone, is able to resist the magnetic
buoyancy \citep{rempel}, \cite{nandi2002} proposed a dynamo model
which restrics for the first time the sunspots' appearance below 45
degrees. In this model, the meridional flow penetrates below the
solar convection zone, where the toroidal magnetic field can't erupt
to the surface, and is dragged in equatorward direction by the meridional
circulation. All these models distinguish for the magnetic
diffusivity and the differential rotation profiles between the radiative
and convective zones, but use a meridional flow based on a single
polytropic approximation to the density profile for both zones.

This work investigates the hypothesis of \cite{nandi2002} of a
meridional return flow in the radiative zone in two other different models.
The first one is just the same model of  \cite{nandi2002}, but with the
magnetic bouyancy mechanism of  \cite{dikchar99}, and is used mainly
as reference considering both confined to convective zone and
tachocline and deep meridional flow penetrating in the radiative zone.
The second one introduces a bipolytropic approximation of the density
profile to construct a meridional flow in the radiative and convective
zones. This density profile better resembles the solar standard model
of \cite{bahcall,bahhcall01},
and it is very interesting to explore how the hypotesis of a deeper
meridional flow works there.
In section 2 we show the mathematical formulation of the problem,
based on the MHD induction equation, and define the meridional flow,
differential rotation, magnetic diffusivity profiles and magnetig
bouyancy mechanism to be used. Section 3 shows the results obtained
with both models. Conclusions and a discussion of our results are
exposed in section 4. Some details of the numerical implementation of
the simulation are shown in the appendix A.

\section[]{Mathematical Formalism}

The MHD induction equation governing the evolution of the magnetic
field is \citep{cowling}

\begin{equation}
\frac{\partial \bf{B}}{\partial t}=\nabla \times (\bf{U}\times
\bf{B})+\eta \nabla^2\bf{B}\label{eq1} \quad. 
\end{equation}
By assuming spherical symmetry, the magnetic and velocity fields can
be writen as
\begin{eqnarray}
\bf{B}&=&B(r,\theta,t) + \nabla \times
(A(r,\theta,t))\label{eq2},\\ 
\bf{U}&=&\bf{u}(r,\theta)+r\sin \theta
\bf{\Omega}(r,\theta)\label{eq3} \quad, 
\end{eqnarray}
where $B(r,\theta,t)$ and $\nabla \times (A(r,\theta,t))$ correspond
to the toroidal and the poloidal components of the magnetic field
respectively; $\Omega$ is the angular velocity,
$\bf{u}=u_r+u_{\theta}$ is the velocity in the meridional plane and
$\eta$ is the magnetic diffusivity.

By replacing equations (\ref{eq2}) and (\ref{eq3}) in
the induction equation (\ref{eq1}) and by separating the poloidal and
toroidal components of the magnetic field, we obtain
\begin{eqnarray}
\frac{\partial A}{\partial t}+\frac{1}{s}(\bf{u}
\cdot\nabla)(sA)=\eta(\nabla^2-\frac{1}{s^2})A +
S_1(r,\theta,t)\label{eq4} \quad ,\\ 
\frac{\partial B}{\partial t}+\frac{1}{r}[\frac{\partial}{\partial
    r}(ru_rB)+\frac{\partial}{\partial
    \theta}(u_{\theta}B)]\label{eq5}=(\bf{B_p} \cdot
\nabla)\Omega\\\nonumber
-\nabla \eta \times \nabla \times
B+\eta(\nabla^2-\frac{1}{s^2})B\nonumber \quad, 
\end{eqnarray}
where $s$$=$$r\sin\theta$ and $\bf{B_p}$$=$$\nabla \times A$. 
It can be observed that
a source term $S_1(r,\theta,B_{\phi})$ has been added (by hand) to the
right side of eq. (\ref{eq4}). This term is very necessary in our 
model for two reasons: it represents the magnetic buoyancy mechanism
which transports the toroidal field from the base of the solar convection
zone to the surface, and it allows the surface regeneration of the
poloidal magnetic field. We will discuss later about the functional
form and physical content of this term.

\subsection{Differential rotation}

As we disscused before, the helioseismology gives a good
characterization of the radial shear and the latitudinal distribution of the
solar angular velocity. It has been found a thin layer of substancial radial
shear called tachocline, located at the base of the solar convection zone,
where the toroidal magnetic field is generated ($\Omega$ effect). An
analytical expression can be inferred from these mesurements, and we
will use the one used by \cite{dikchar99}, who were the first to
include a solar-like differential rotation profile in a kinematic
model, 
\begin{equation}
\Omega(r,\theta)=\Omega_c+\frac{1}{2}[1+erf(2\frac{r-r_c}{d_1})](\Omega_s(\theta)-\Omega_c)
\quad .
\label{eq6}   
\end{equation}
Here, $\Omega_s(\theta)$$=$$\Omega_{Eq}+a_2\cos^2\theta+a_4\cos^4\theta$
is the latitudinal differential rotation in the surface and $erf(x)$ is
an error function that confines the radial shear to a tachocline of
thickness $d_1$$=$$0.05 \sradi$. In this expression, a rigid core rotates uniformly with
angular velocity $\Omega_c/2\pi$$=$$432.8$. Other values are
$\Omega_{Eq}/2\pi$$=$$460.7$, $a_2/2\pi$$=$$-62.9$,
$a_4/2\pi$$=$$-67.13$ nHz and $r_c$$=$$0.7\sradi$.

\subsection{Meridional circulation}

An analytical expression for the velocity in the meridional plane is
more difficult to find that the previous one, because there is not 
enough observational data. On one hand,
the observations suggest a poleward flow with an average velocity of
$10$$-$$20$ m/s \citep{giles97} in the solar surface, but it is little
what we know about the return flow, except that it must exist to
satisfy mass conservation.  
On the other hand, numerical simulations of turbulent convection zones
show that a structure of plumes in the base of the solar
convection zone is able to push down the flow (and the magnetic field)
to the radiative zone, where
there is a net movement towards the equator, with a mean velocity of
around $3$ m s$^{-1}$ \citep{brumell,miesch}. In this sense, we consider a
single convection cell for each 
meridional quadrant, based on the next equation introduced by
\cite{dikpati94,choudhuri95} and previosly used by \cite{nandi2002}:
\begin{equation}
\rho(r) \bf{u}=\nabla \times [\psi(r,\theta)e_{\phi}]\label{eq7} \quad,
\end{equation}
where $\psi$ is the stream function given by
\begin{eqnarray}
\psi r\sin\theta &=& (r-R_b)\psi_0
\sin[\frac{\pi(r-R_b)}{(\sradi-Rb)}]\\\nonumber
&\times&(1-e^{-\beta_1r\theta^{\epsilon}})(1-e^{\beta_2r(\theta-\pi/2)})\\\nonumber
&\times&e^{[(r-ro)/\Gamma]^2}\label{eq8} \quad,   
\end{eqnarray}
and $\rho$ is the density profile for the sun. Most of the 
models use a profile for an adiabatic gaseous sphere with a
specific heat ratio coefficient $\gamma$$=$$5/3$, which corresponds to a
constant polytropic index $m$$=$$1.5$. So we have
\begin{equation}
\rho(r)=C(\frac{\sradi}{r}-0.95)^m\label{eq9} \quad ,
\end{equation}
where we chose $C$$=$$3.60 \times 10^{-3}$ gr cm$^{-3}$ as the surface
density value \citep{gpinzon}. The coefficient $\psi_0$ was chosen in
such a way that the 
maximal latitudinal velocity at middle latitudes is $20$ m s$^{-1}$. Other
values in the eq. (\ref{eq8}) are: $\beta_1$$=$$1.65 \times 10^{10}$ cm$^{-1}$,
$\beta_2$$=$$2.2\times10^{10}$ cm$^{-1}$, $\epsilon$$=$$2.0000001$,
$r_o$$=$$(\sradi-R_{min})/4.15$ and $\Gamma$$=$$3.47\times
10^{10}$ cm. Here $R_{min}$ is the minimal $r$ coordinate value for the
integration range, and the free parameter $R_b$ is the maximal depth
of return flow (See  \cite{dikpati94} for more details).

\subsection{Magnetic buoyancy (MB)}

We mentioned above that a buoyancy magnetic force upwards is
generated in the base (or below) of the solar convection zone, due to
of the density difference between the inner and the outer regions of a
magnetic flux tube. Simulations in this sense have shown that toroidal
magnetic tubes of $10^5$ G emerge to the surface and are twisted by
Corilis forces to form tilted bipolar active magnetic
regions (tilted BMRs). They have also shown that the tilt and the
latitud of those BMRs 
strongly depend of the initial value of the toroidal magnetic
field. Initial values greater than $1.6\times10^5$ G don't produce
tilt in BMR, and values lower than $6\times 10^4$ emerge radially with
tilts in disagreement with Joy's law
\citep{dsilva93,fan93,caligari95,calig98}. We introduce a simplified
form of including these results, due to \cite{dikchar99}, with an
initial value of  $10^5$ G. Its analytical form is
\begin{eqnarray}
  S_1(r,\theta;B)&=&\frac{S_o}{4} 
  B(r_c,\theta,t)[1+erf(\frac{r-r_2}{d_2})]\label{eq10}\\\nonumber&\times&[1-erf(\frac{r-r_3}{d_3})]\\\nonumber&\times&[1+(\frac{B(r_c,\theta,t)}{B_0})^2]^{-1}
  \cos\theta\quad.
\end{eqnarray} 

It can be seen that this term acts non-locally in $B$ \citep{d95,d96,d97}:
values of 
toroidal magnetic field $B$ at the tachocline produce proportional poloidal
magnetic fields at a thin layer close to the surface, defined by the
error (erf) functions.
The last term, $[1+(\frac{B(r_c,\theta,t)}{B_0})^2]^{-1}$, anihillates
the subsequent increase of $A$ at the surface, beyond some maximal
level $B_0$. It is, in other words, a saturation term, and it is the
only source of non-linearity in the system. The other parameters are
$r_2$$=$$0.95$$\sradi$, $r_3$$=$$\sradi$,
$d_2$$=$$d_3$$=$$0.025$$\sradi$, so the alpha effect is confined
beneath the surface in a layer of thickness $0.025$$\sradi$.

\subsection{Magnetic diffusivity}

Magnetic diffusivity $\eta$ is different for the radiative and the convective
zones. The turbulent regime present at the convective zone makes
$\eta$ two orders of magnitude higher than the radiative one (see
fig 1 of \cite{dikchar99}),
\begin{equation}
\eta(r)=\eta_c + \frac{\eta_T}{2}[1+erf(2\frac{r-r_c}{d_1})] \quad,
\label{eq11}
\end{equation}
with  $\eta_c$$=$$2.2\times 10^9$ and $\eta_T$$=$$0.5\times 10^{11}$ cm$^2$
s$^{-1}$ and $r_c$$=$$0.7$$\sradi$. 

\section{Results}
We solved the  model described above by means of the ADI method on a
two-dimensional mesh of $64$$\times$$64$ spatial divitions, with $0.55
\sradi \le r \le 1 \sradi$ and $0 \le \theta \le \pi/2$, and by
imposing the following boundary conditions:
\begin{eqnarray}
\text{at  } &\theta=0:&A=0\text{, }B=0 \quad,\label{eq12}\\
\text{at  } &\theta=\pi/2:& \frac{\partial A}{\partial\theta}=0
\text{,  } B=0 \quad,\label{eq13}\\
\text{at  }&r=0.55\sradi:&A=0\text{, }B=0\quad,\label{eq14}\\
\text{at  }&r=\sradi:&B=0\quad\label{eq15}.
\end{eqnarray}
At the upper boundary it must be ensured, in addition, that $A$
satisfies the free space condition, 
\begin{equation}
(\nabla^2-\frac{1}{r^2 \sin^2 \theta})A=0\quad.\label{eq16}
\end{equation}
The initial condition is not determinant for the evolution of the model,
as long as the $S_0$ coefficient is large enough to avoid the magnetic
field to diffuse. In this case the system always relaxes to peridic
solutions. In all cases, we took the initial condition
\begin{equation}
B(r,\theta,0)=B_0 \sin (2\theta) \quad.\label{eq17}
\end{equation}
The code was tested by using all profiles and parameters of
\cite{dikchar99}, and by reproducing the results therein.
The numerical details on discretization and integration are shown in
appendix A.

\subsection{Meridional flow confined to the convective zone}
\begin{figure}
  \includegraphics{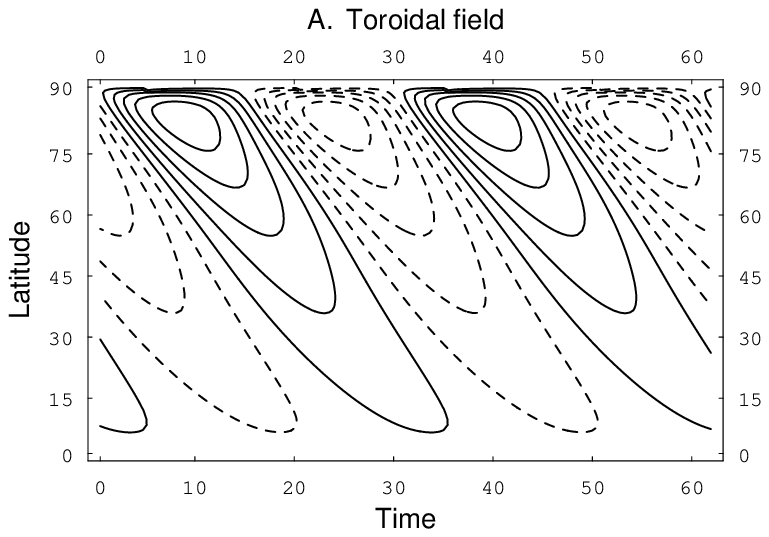}
  \includegraphics{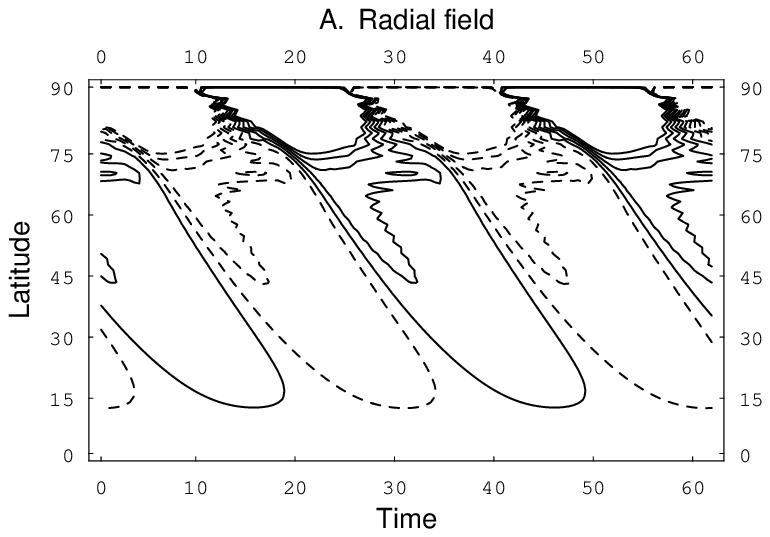}
  \caption{Butterfly diagrams for the, A, toroidal magnetic field at
  the base of  solar convection zone ($r$$=$$0.7\sradi$) and, B, radial
  field at the surface ($r$$=$$\sradi$), by using the parameters shown
  in Table \ref{table1}. The contours are equally spacied, with solid (dashed)
  lines for positive (negative) values. Time in years and latitude
  in degrees.}
  \label{fig1}
\end{figure}

In this first model, the meridional flow penetrates until $0.675 \sradi$,
that is, it is confined to the solar convection zone and the tachocline. The
free parameters used in this model are shown on the left side of
Table \ref{table1}. As the butterfly diagram shows (Fig
\ref{fig1}A), the maximal toroidal magnetic field in the base of SCZ is
located at latitudes between $75$ and $90$ degrees, with weak branches
migrating in the equatorward direction. Under the primordial assumption
that the toroidal field lines emerge to the surface like magnetic flux
tubes,  this model generates sunspots close to the poles. The
rigth side of Table \ref{table1} summarizes the most important results of this
model. We obtain a period slightly larger than the one observed and
the maximal intensities of the magnetic fields are in the correct order
of magnitude for the toroidal one but one order the magnitude over
the one expected for the radial field.
\begin{table}
  \begin{center}
    \caption{Free parameter values for a meridional flow confined to
    the convective zone} 
    \begin{tabular}{c|c|c|c}\hline 
	Parameter & Value & Result & Value\\ \hline
	$U_0$ & $2000$ cm s$^{-1}$ & $B_{\phi}$max & $2.83 \times 10^5$ G \\ 
	$S_0$ & $5$ cm s$^{-1}$ & $B_r${max} & $150$ G \\ 
	$\eta_T$ & $0.8 \times 10^{11}$ cm$^2$ s$^{-1}$ & T & $27.8$ years\\ 
	$R_b$ & $0.675 \sradi$ \\\hline
	\label{table1}
     \end{tabular}
  \end{center}
\end{table}
\subsection{Deep meridional flow}
A different situation is observed if the
meridional flow penetrates into the radiative zone, as
deep as $r$$=$$0.61\sradi$. A better distribution of the
toroidal lines is obtained, however, a polar branch still remains, in
contrast to \cite{nandi2002}. There are two zones of maximal
intensity: one below $45$ degrees and another above $80$ degrees (at
the poles). This result can be interpreted in the following way: 
if most of the magnetic field is dragged below the tachocline,
only a small portion emerges to the surface by the action
of the magnetic bouyancy force.
The rest migrates to the equator, dragged by
the solar plasma, and will undergo the alpha effect only at
middle and low latitudes.

It is important to notice that this model gives
a period of $28.8$ years (close to the observed one) and maximal values of both
toroidal and radial fields within the right order of magnitudes. The  
parameters we used and the results are summarized in 
Table \ref{table2}. The large scale behaviour of the model
can be observed in the butterfly diagrams of  Fig 2.
The equatorward migration of the branches of toroidal magnetic field
is evident.
\begin{table}
  \begin{center}
    \caption{Free parameter values and results for a deep meridional flow} 
    \begin{tabular}{c|c|c|c}\hline 
	Parameter & Value & Result & Value\\ \hline
	$U_0$ & $2000$ cm s$^{-1}$ & $B_{\phi}$max & $2.58 \times 10^5$ G \\ 
	$S_0$ & $5$ cm s$^{-1}$ & $B_r${max} & $50$ G \\ 
	$\eta_T$ & $0.8 \times 10^{11}$ cm$^2$ s$^{-1}$ & T & $28.8$ years\\ 
	$R_b$ & $0.61 \sradi$ \\\hline
     \end{tabular}
  \label{table2}
  \end{center}
\end{table}
\begin{figure}
  \includegraphics{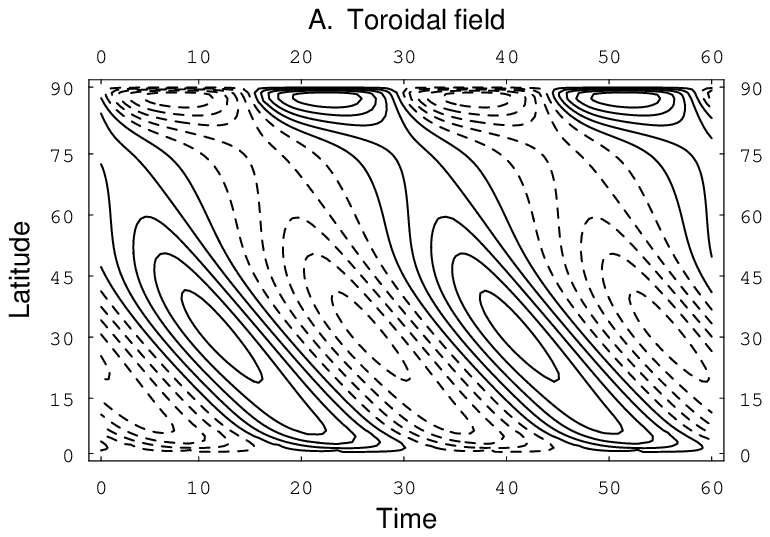}
  \includegraphics{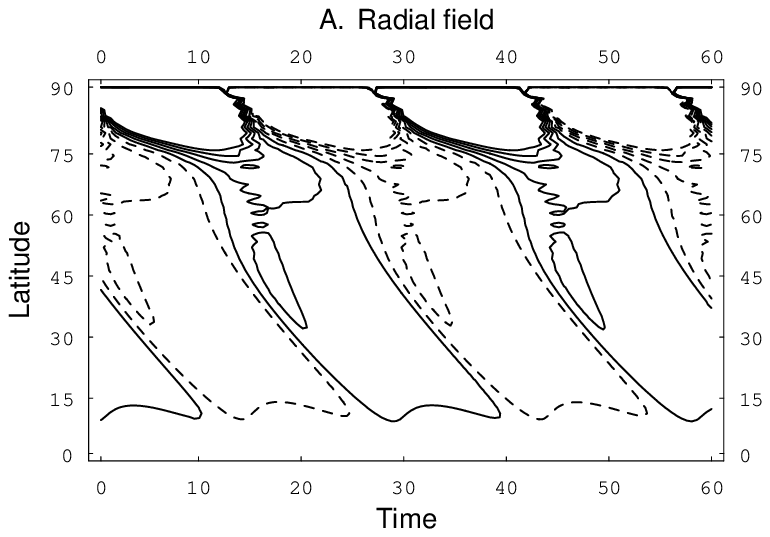}
  \caption{Butterfly diagrams for the, A, toroidal magnetic field in
  the base of  solar convection zone ($r$$=$$0.7\sradi$) and, B, radial
  field in the surface ($r$$=$$\sradi$) with the parameters showed in table 
  \ref{table2}. Time in years and latitude in degrees.}
  \label{fig2}
\end{figure}
\subsection{A more realistic density profile}
\begin{figure}
  \includegraphics{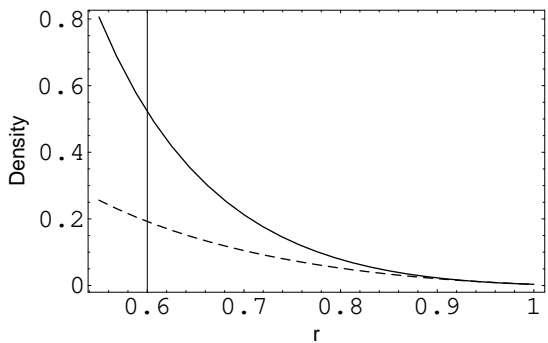}
  \caption{Density profiles in the solar interior for the usual way
  (dashed line) and the bipolytropic model (solid line). In both
  graphics the density has been normalized to a surface value of
  $3.603 \times 10^{-3}$g cm$^{-3}$ like the one in the bipolytropic
  profile. Density in g cm$^{-3}$, $r$ in $\sradi$.} 
  \label{fig3}
\end{figure}

As the model of Nandy and Choudhouri does \citep{nandi2002}, the
model above employs a single polytropic density profile
(Eq. \ref{eq9}), with $m$$=$$1.5$ ($\gamma$$=$$5/3$). 
This is just true for the convective zone ($r$$>$$0.7$), 
but not for the radiative one.
We can introduce a more realistic density profile by using a 
bipolytropic aproximation due to \cite{gpinzon},
with $\gamma$$=$$5/3$ for the convective zone and $\gamma$$=$$1.26$
for both radiative zone and core (Fig \ref{fig3}). We fit the
numerical results of \cite{gpinzon} by the analytical expression
\begin{equation}
\rho(r)=(\frac{\sradi}{r}-0.904)^{2.4}\label{rho_poli},
\end{equation}
with the same surface density value as (\ref{eq9}).
\begin{table}
  \begin{center}
    \caption{Free parameter values and results for a more realistic
    density profile model} 
    \begin{tabular}{c|c|c|c}\hline 
	Parameter & Value & Result & Value\\ \hline
	$U_0$ & $2000$ cm s$^{-1}$ & $B_{\phi}$max & $3.77 \times 10^5$ G \\ 
	$S_0$ & $5$ cm s$^{-1}$ & $B_r${max} & $300$ G \\ 
	$\eta_T$ & $0.8 \times 10^{11}$ cm$^2$ s$^{-1}$ & T & $72.1$ years\\ 
	$R_b$ & $0.61 \sradi$ \\\hline
	\label{table3}
     \end{tabular}
  \end{center}
\end{table}
\begin{figure}
  \includegraphics{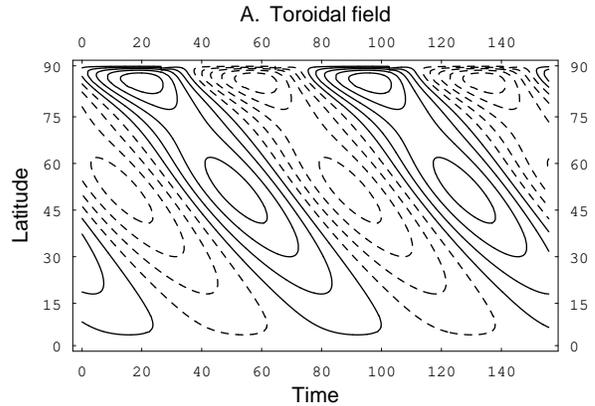} 
  \caption{Butterfly diagrams for the toroidal magnetic field in
  the base of  solar convection zone ($r$$=$$0.7\sradi$) for a
  meridional flow penetrating into the radiative zone and a
  bipolytropic density profile.}
  \label{fig4}
\end{figure}

Despite that this profile closely resembles the internal
structure of the solar standard model of \cite{bahcall,bahcall01}, our
results are not so good as before.

On one hand, we obtain again two maximal intensity
zones: the first one very close to the poles, but the second one at
middle latitudes, between $60$ and $40$ degrees, in disagreement with the
observations. 
On the other hand, since the bipolytropic density profile is
around three times larger that the usual density profile at the return point,
$r$$=$$0.61$$\sradi$, (see Fig \ref{fig3}),
the counter flow is around three times lower 
that before in order to hold mass conservation, $\nabla \cdot (\rho{\bf
  U})$$=$$0$. 
Thus, our model gives a new period much larger than 
observed (actually, three times larger, but this is not a linear
relationship). If we allow the flow to go deeper, the second zone 
of maximum toroidal field approaches the equator, but at the cost of
increasing the period very quickly.

\section{Conclusions}
In this work three different models of solar dynamo are shown.
We developed these models by using the velocity field
of \cite{nandi2002}, the magnetic buoyancy mechanism
introduced by \cite{dikchar99} and different values of the
diffusivity coefficient for the convective and the radiative
zones, as in \cite{dikchar99,nandi2002}. In other words we have
recovered the \cite{nandi2002} results even with a different source  
formulation. This suggests that the good performace of the kinematic models in
the papers above is general in nature. 
Moreover, our results suggests that, at least for this kind of models, the
effect of a deep meridional flow to solve the apperance of sunspots at high
latitudes is general, too.  

The most important results are:

\begin{enumerate}
\item If the meridional flow is confined to the convective zone ($r$$>
  $$0.675\sradi$), the emergence latitude of sunspots is just near 
  the poles, as expected. This model also gives wrong maximal values
  of the radial field intensities. 
\item When the flow penetrates until $0.61\sradi$, the model gives
  a better solution. However, two regions of maximal toroidal field appear (in
  contrast with the single region obtained by \cite{nandi2002}):
  one of them is located in the right range of latitudes and the
  other remains very near to the poles. Except for this last region,
  the model approximates well the observed butterfly diagram. 
\item In the third model we attempt to improve it by
  incorporating  a more realistic density profile, but the results are
  not satisfactory. There are two zones 
  of maximal toroidal field as before, but the second one 
  appears at latitudes to high to agree with observations. Moreover,
  the period of cycle is three times larger than observed.
\end{enumerate}

Our simulations show again the good effects of a deep meridional flow to solve
the appearance of sunspots at high latitudes, proposed by  \cite{nandi2002}, 
but this hypothesis deserves a more extended discussion.
First, recent results on magnetic tube flux simulations \citep{rempel}
suggests that a subadiabatic stratification may suppress the magnetic
buoyancy force and address the equilibrium of a toroidal band. Despite that
\cite{rempel} do not speak of a deep flow, such kind of stratification is the
one to be found in the radiative zone and the tachocline, supporting the
process suggested by \cite{nandi2002} to drag the toridal field towards lower
latitudes. 
However, they do not use a subadiabatic stratification, but an
adiabatic background ($\gamma$$=$$5/3$) everywhere. 
We have introduced a subadiabatic background in a simple way, by using a
more realistic solar structure with a
bipolytropic density profile, but our results are not in
agreement with the observations. Nevertheless, this may suggest that a more
realistic meridional circulation profile is needed, not that the idea of a
magnetic buoyancy suppressed by a sub-adiabatic stratification should be
rejected. 

Second, it should be noticed that the discussion around a deep
meridional flow hipothesis is not closed. On one hand, turbulent
convection models obtain flow penetration (in a plumes dominated
structure) in the stable layer \citep{miesch,tobias01}, and some
analysis on helioseismological observations give a poleward meridional
flow across the entire convective zone \citep{giles97}, supporting the
idea of a deep flow.   
On the other hand, the meridional flow is such a weak flow that it is
very  unlikely that it can penetrate the strongly subadiabatically 
stratified radiative core of the Sun.
In addition, other problems would have to be reconsidered if a deep
flow is assumed, like a larger angular momentum transfer to the
radiative zone \citep{durney} and the changes in the relative
abundance of Lythium and other elements in both radiative and
convective zones \citep{zahn92,brun99}, but these are still open
problems. Despite these uncertainties, our aim with this paper is to
verify what would happen if this scenario were possible.

\section*{Acknowledgments}
We thank G. Pinzon for his colaboration with the bipolytropic density
data, P.D. Mininni and D. Nandy for help and useful
discussions, and an annonymous referee for his invaluable and accurate
suggestions.  This work was supported by the Bogota Research Division
of the National University of Colombia, project DIB-803755.

\vspace{1cm}

\appendix
\section[]{Numerical methods}
The equations (\ref{eq4}) and (\ref{eq5}) are two
coupled partial differential equations of advection-diffusion type,
with a non-linear term ($S_1$). There are two important aspects to have
into account: numerical dicretization and numerical integration. In the
first one, the advective-diffusive character of the equations must
be considered. In order to ensure precision and stability
for all times the advective terms are discretized by using the Lax-Wendroff
mechanisms, which is a second-order discretization in both space and
time, and the diffusive terms are discretized by a simple forward-time
centered-space (FTCS) mechanism \citep{ames,press}. 
In the second one, the two dimmensional character 
and the non-linearity are considered. Alternating Direction-Implicit
(ADI) method is a good numerical technique for these
characteristics \citep{ames,press}.

A first step is to write the equations in operator notation, as follows:   
\begin{eqnarray}
  \frac{\partial A}{\partial t}=(L_r + L_{\theta})A + S_1\label{op1} \quad,\\
\frac{\partial B}{\partial t}=(L_r + L_{\theta})B + S_2\label{op2}
\quad ,
\end{eqnarray}
where
\begin{eqnarray}
L_rA&=&-\frac{u_r}{r}A-u_r\frac{\partial A}{\partial r}-\frac{\eta}{2
  r^2 \sin^2 \theta}A\\\nonumber &+&\frac{2 \eta}{r}\frac{\partial A}{\partial
  r}+\eta\frac{\partial^2 A}{\partial r^2}\label{opa1}\quad , 
\end{eqnarray}
\begin{eqnarray}
L_{\theta}A&=&-\frac{1}{r}\cot\theta u_{\theta}A -
  \frac{u_{\theta}}{r}\frac{\partial A}{\partial
  \theta}-\frac{\eta}{2r^2 \sin^2 \theta}A \\\nonumber &+& \frac{\eta}{r^2}\cot
  \theta \frac{\partial A}{\partial \theta} + \frac{\eta}{r^2}
  \frac{\partial^2 A}{\partial \theta^2}\label{opa2}\quad ,\\
L_rB&=&-u_r\frac{\partial B}{\partial r}-\frac{u_r}{r}B-\frac{\partial
  u_r}{\partial r}B - \frac{\eta}{2r^2 \sin^2 \theta}B \\\nonumber &+& \frac{2
  \eta}{r}\frac{\partial B}{\partial r} + \eta \frac{\partial^2
  B}{\partial r^2}+\frac{\partial \eta}{\partial r}(\frac{B}{r}+\frac{\partial
  B}{\partial r})\label{opb1} \quad ,\\ 
L_{\theta}B&=&-\frac{1}{r}\frac{\partial u_{\theta}}{\partial \theta}B
  -\frac{u_{\theta}}{r} \frac{\partial B}{\partial
  \theta}-\frac{\eta}{2r^2\sin^2\theta}B \\\nonumber &+&
  \frac{\eta}{r^2}\cot\theta 
  \frac{\partial B}{\partial \theta} +
  \frac{\eta}{r^2}\frac{\partial^2B}{\partial \theta^2}\label{opb2}
  \quad ,
\end{eqnarray}

and $S_1$ and $S_2$ are the crossed terms. $S_1$ is
given by equation (\ref{eq10}) and $S_2$ is given by
\begin{equation}
r*\sin\theta (B_p \cdot \nabla)\Omega=\frac{1}{r} \frac{\partial
  \Omega}{\partial \theta} \sin\theta (-A-r\frac{\partial A}{\partial
  r}) + \frac{\partial \Omega}{\partial r}(A\cot\theta +
\frac{\partial A}{\partial \theta}). 
\end{equation}

In the ADI method, the time step is divided into two steps of size
$\Delta t/2$. In each half step, one spatial dimmension is treated
implicitly and the other is treated explicitly. We treated 
the $\theta$ terms in implicit form and the $r$ terms in
explicit form in the first half step,  as follows:

\begin{eqnarray}
\frac{A_{ij}^{n+1/2}-A_{ij}^n}{\Delta
  t/2}&=&L_{\theta}A^{n+1/2}+L_rA^n + S_1^n ,\\ 
\frac{B_{ij}^{n+1/2}-B_{ij}^n}{\Delta
  t/2}&=&L_{\theta}B^{n+1/2}+L_rB^n + S_2^n ,
\end{eqnarray}
where $i,j=0,1, ...,N$ are the spatial divisions and $n$ is the time
step. The next half step is treated in the reverse manner,
\begin{eqnarray}
\frac{A_{ij}^{n+1}-A_{ij}^{n+1/2}}{\Delta
  t/2}&=&L_{\theta}A^{n+1/2}+L_rA^{n+1} + S_1^{n+1/2},\\ 
\frac{B_{ij}^{n+1}-B_{ij}^{n+1/2}}{\Delta
  t/2}&=&L_{\theta}B^{n+1/2}+L_rB^{n+1} + S_2^{n+1/2}.
\end{eqnarray}

The source terms are always treated explicitly, to guarantee the
linearity of equations in $A$ and $B$. These equations can be
organized in such a way that they can be solved by a standard
tridiagonal algorithm. The numerical treatment of the boundary
conditions is developed in detail in \cite{dikpati95}.

\bsp

\label{lastpage}

\end{document}